\documentclass[sigconf,nonacm]{acmart}

\acmConference[ICSE 2024]{46th International Conference on Software Engineering}{April 2024}{Lisbon, Portugal}

\AtBeginDocument{%
  \providecommand\BibTeX{{%
    \normalfont B\kern-0.5em{\scshape i\kern-0.25em b}\kern-0.8em\TeX}}}

\begin{document}

\title{Code Ownership in Open-Source AI Software Security}

\author{Jiawen Wen$^{1}$, Dong Yuan$^{1}$, Lei Ma$^{2,3}$, Huaming Chen$^{1}$}
\affiliation{%
  \institution{$^1${The University of Sydney, Australia}\\ \quad $^2${University of Alberta, Canada} \\ \quad $^3${The University of Tokyo, Japan}
  \country{} 
}
}

\renewcommand{\shortauthors}{Jiawen et al.}

\begin{abstract}
    As open-source AI software projects become an integral component in the AI software development, it is critical to develop a novel methods to ensure and measure the security of the open-source projects for developers. Code ownership, pivotal in the evolution of such projects, offers insights into developer engagement and potential vulnerabilities. In this paper, we leverage the code ownership metrics to empirically investigate the correlation with the latent vulnerabilities across five prominent open-source AI software projects. The findings from the large-scale empirical study suggest a positive relationship between high-level ownership (characterised by a limited number of minor contributors) and a decrease in vulnerabilities. Furthermore, we innovatively introduce the time metrics, anchored on the project's duration, individual source code file timelines, and the count of impacted releases. These metrics adeptly categorise distinct phases of open-source AI software projects and their respective vulnerability intensities. With these novel code ownership metrics, we have implemented a Python-based command-line application to aid project curators and quality assurance professionals in evaluating and benchmarking their on-site projects. We anticipate this work will embark a continuous research development for securing and measuring open-source AI project security.
\end{abstract}

\ccsdesc[300]{Software and its engineering~Risk management}

\keywords{Open-Source Software, AI, Security Management, Code Ownership, Process Metrics, Empirical Software Engineering}



\maketitle

\section{Introduction}

In the trending open-source AI software, where platforms such as GitHub, TensorFlow, PyTorch, and OpenCV are prevalent, software vulnerabilities have become a paramount concern. This concern is especially significant with the emergence of groundbreaking projects like ChatGPT, which mark a new era in AI. It is intensified by the transparent nature and the anonymity of contributors in open-source projects, leading to a heightened risk of unforeseen threats, as outlined in studies of \cite{ladisa2023,enisa2021}, and highlighted in the US Executive Order on Advancing National Cyber Security \cite{biden2021}. These studies underscore the importance of developer activity as a vector for attack research and mitigation, emphasizing how attackers can infiltrate systems by posing as maintainers or contributors. 

In this context, code ownership becomes a crucial metric for evaluating developer involvement and identifying latent vulnerabilities in the AI software projects. The correlation between code ownership and software quality indicates that understanding developer dynamics through code ownership metrics can assist in revealing software vulnerabilities and potential software failures~\cite{bird2011donttouch,greiler2015ownership}. This multifaceted approach, which combines insights into developer activity, code ownership, and the unique characteristics of open-source AI software, is essential for developing a comprehensive security framework. It further extends the capability of adapting to the challenges posed by the open-source AI software supply chain.

This leads to the pivotal question: `\textit{To what extent does code ownership influence and reflect the security and vulnerability of open-source AI software?}' A clear understanding of this aspect equips project managers, maintenance teams, and quality assurance engineers with necessary tools to improve project governance, identify security issues, and enhance user protection. Thus, in this work, we explore the nexus between code ownership and software vulnerability in open-source AI projects, presenting following contributions:

\textbf{1. Novel code ownership metric:} We propose a novel code ownership metric tailored for open-source AI application security. The proposed code ownership metric integrates software component frequency/proportion and time/release attributes, offering deeper insights into the link between code ownership and vulnerabilities in open-source AI software.

\textbf{2. Quantitative analysis:} We conduct a rigorous quantitative analysis dissecting the interplay among the code ownership metric, temporal factors, and software iterations across five open-source AI projects with vulnerabilities. The results show the effectiveness and trustworthiness of the proposed code ownership metric.

\textbf{3. Comparative analysis:} We extend the work with a comparative analysis juxtaposing the new metric with the origin code ownership metrics and classic metrics. By scrutinizing potential bias-inducing variables, a deeper understanding of this bespoke metric in the open-source AI software realm is achieved.

\textbf{4. Tool Development:} We implement a Python-based command-line tool\footnote{https://github.com/jemjemzzZ/Code-Ownership}, to assist developers and quality assurance experts in computing both repository-wide and file-specific code ownership.

\section{Related Work}
In this section, we discuss existing literature works outlining the impacts of developer contribution practices on software quality and security, which is primarily on traditional software projects. Focusing on Windows Vista and Windows 7, Bird et al.~\cite{bird2011donttouch} devised an ownership calculation based on the volume and distribution of commits for each software component. The basic software component is defined in operating systems, abundant in executable and driver files like \textit{.dll}, as a compiled binary. The distinct connection between the ratio of minor contributors and the failure rates of software components was analysed. The link between minor contributors and pre-release software failures is notably strong, with significance larger than 85\%. However, this correlation diminishes for the post-release. For metrics of major contributors and ownership index, an association with software failures is not as evident. 

Moreover, other studies identified a relationship between basic software characteristics (i.e., \textit{size}, \textit{complexity}, \textit{churn rate}) and the software bugs~\cite{nagappan2005churn, ostrand2004bugs}. Some studies even argue that omitting \textit{size} as a factor might significantly affect certain code metrics~\cite{elemam2001confounding}. \cite{bird2011donttouch} also compares indicators of minor contributor with classic code metrics in terms of software defects. During the pre-release phase, minor contributors demonstrated a stronger correlation of 10\%-20\%. However, in the post-release, these contribution metrics did not exhibit a stronger correlation with software failures compared to classic metrics such as \textit{size}, \textit{churn}, and \textit{complexity}. When comparing the code ownership metric to classic metrics (base metric), the minor contributor metric exhibited stronger correlation by improving the result by about 20\%. Conversely, indicators related to major contributor and ownership index did not contribute substantially and, in some scenarios, were even counterproductive.

Though \cite{bird2011donttouch} was criticised for the unclear definition and selection logic of the crucial `\textit{software component}' as a key concept for their analysis, Foucault et al. \cite{foucault2014ownership} extended the work in a replication study. They considered Java FLOSS projects as the reseaerch subject, and introduced two additional levels for software component granularity: \textit{Java packages} and \textit{source files}. They observed that the code ownership concepts in Bird et al.'s work had limitations to Java FLOSS projects, exhibiting correlations significantly lower than those in Bird et al.'s study. Foucault et al. identified a stronger connection with ownership laws at the package level compared to the file level, particularly in the latest release as opposed to  throughout the entire release history. They also observed differences in contributor dynamics between industrial and FLOSS projects, noting that FLOSS projects tended to feature a few major contributors alongside many minor contributors, affecting the impacts of minor contributors on ownership metrics.

Greiler et al. \cite{greiler2015ownership} conducted a study within the Microsoft environment, focusing on Office and Windows, and introduced `code directories' as a new granularity for software components. They found a predominant trend of having one or two contributors per file or directory, similar to the pattern in FLOSS projects reported by Foucault et al. However, Greiler et al. deviated from the conclusions by Foucault et al., showing a significant correlation between code ownership and software failures. This underscores the significance of code ownership metrics at both the file and directory levels.

Recently, Ladisa et al. delineate various types of open source software supply chain attacks, highlighting that developer actions are a notable avenue for both identification and exploit of attacks~\cite{ladisa2023}. According to their classification of attacks targeting open-source software supply chains, infiltrating as a maintainer or contributor and imitating typical developer behavior can be identified as initial steps for different attack strategies. These attacks frequently involve subtle and hard-to-detect code changes, including the injection of harmful dependencies, creating name confusion with authentic packages, and employing other techniques.

\begin{figure}
    \centering
    \includegraphics[width=\linewidth]{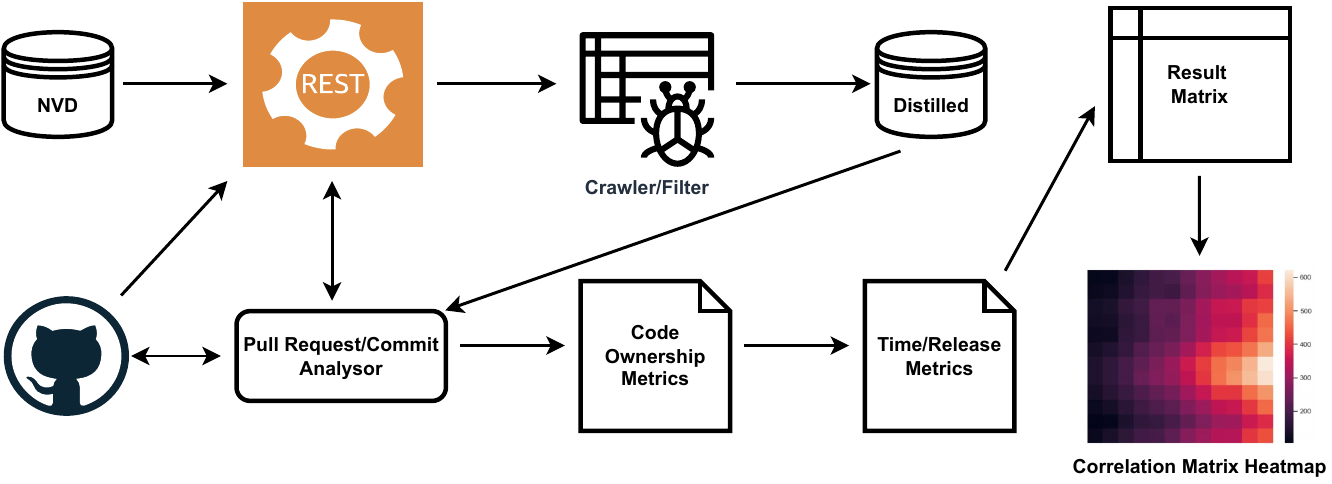}
    \caption{The overall experiment process}
    \label{fig:process}
\end{figure}

In this work, we leverage the existing works and extend the focus to open-source AI software projects. The main objective is to investigate the distinct features of open-source AI software and incorporate code ownership metrics to formulate appropriate definitions for software component. This process aids in developing effective vulnerability classification techniques. We thoroughly design our experiment following the three phases depicted in Figure \ref{fig:process}, covering data collection and process, metrics computing, and correlation matrix heatmap analysis, to evaluate the impact of code ownership metrics on open-source AI software.

\section{Research Questions and Hypotheses}
\subsection{Research Questions}

Our work primarily centers on the formulation and effectiveness investigation of code ownership metrics within open-source AI software projects. It explores their relationship with software vulnerabilities, and examines their applicability throughout different stages of software development in practice. To achieve the research objectives, the work poses following research questions:

\textbf{RQ1.} How are code ownership metrics developed in open-source AI projects, and how do these metrics correlate with the characterization of software components and vulnerabilities?
    
\textbf{RQ2.} How does the proposed metric for analyzing open-source AI projects fare in terms of effectiveness, accuracy, and resilience when compared to Bird et al.'s original code ownership metric and classic process metrics? Additionally, what is the impact on the categorization of vulnerability severity when variables and thresholds related to code ownership are modified?

\textbf{RQ3.} Does the proposed code ownership metric change with the development stage, and what practical uses can be derived from it?

\subsection{Hypotheses}

Whilst analysing the available GitHub repositories for various open-source AI projects, we note that the expertise of minor developers may varies significantly. This suggests that an increase in the number of minor contributors could lead to a heightened risk of vulnerabilities within software components. In the meantime, the study posits that software components, namely source code files, maintain relative independence in terms of code ownership, lacking a discernible pattern. Thus, we propose three hypotheses:

\textbf{Hypothesis 1.} \textit{The vulnerability of software components increases with a higher number of minor contributors, which is influenced by the duration and operational practices of the project's development stage.} This hypothesis derives from \cite{bird2011donttouch}, suggesting that the likelihood of software failures increases as the absolute number of minor contributors grows. Additionally, the proportion of minor contributors, reflective of wider participation in the project, exacerbates this effect. The operational practices and the maturation of the project also contribute to this dynamic, indicating that as a project progresses, the increasing number of minor contributors, both in absolute terms and as a share of total contributions, may lead to a higher incidence of vulnerabilities. This underscores the importance of considering both the quantity and the relative influence of minor contributors when evaluating the robustness of software components throughout the project lifecycle.

\textbf{Hypothesis 2.} \textit{The vulnerability susceptibility of a software component is unrelated to its vulnerability occurrence rate.} While previous works \cite{crnkovic2011classification, lau2005taxonomy, meneely2013patch, arafat2009commit} suggest patterns of vulnerability commit rates associated with specific sections of source code, our observations 
primarily revolve around the software component in isolation, rather than a holistic review. In terms of code ownership, our analysis delves into the commit history of individual source code files, giving equal weight to each commit.

\textbf{Hypothesis 3.} \textit{The software component's behavior is independent of its location within the scope of the project.} Concerning intra-project dependencies, extensive researches underscore dependencies as a prominent vector for introducing vulnerabilities into open-source software. However, the underlying rationale often suggests that vulnerabilities arise from external sources, such as outdated libraries or malicious dependency injections \cite{plate2015impact, pashchenko2018vulnerable, kula2018developers, mostafa2017experience}. This implies that dependencies intrinsic to the software project itself appear to exert minimal influence on the vulnerabilities.

\section{Terminology and Metrics}

\subsection{Terminology}

To address the research questions, we introduce several ownership metrics to assess the hypotheses and gauge the performance of ownership. Following we discuss some crucial terms in this work.

\textbf{Software Component.} Software component represents the foundational unit of a software project, encompassing its core functionalities where the defects can be traced back to. In this work, which concentrates on open-source software, the `\textit{software component}' refers to a source code file.

\textbf{Contributor.} Contributor is the individual who has manipulated the software component with commits or software alterations.

\textbf{Contribution.} Contribution is denoted by successful commits or changes. Each such commit is regarded as a singular contribution.

\textbf{Proportion of Ownership.} The proportion of ownership (simply called `\textit{Ownership}') of a contributor for a particular software component calculates the contributor's involvement in a software component by determining the ratio of contributions made by the individual against the total contributions to that component.

\textbf{Minor Contributor.} Minor contributor (`\textit{Minor}') is the developer who, despite contributing to a component, holds an ownership percentage below a predetermined threshold T\%. This research sets the threshold at 10\%, as derived from the experiment (see Sec.~\ref{cha:distortion_check}).

\textbf{Time Stage.} Time stage pertains to the various development phases of an open-source AI software project, categorized into five distinct stages based on the project's lifespan.

\textbf{OSS Stage.} OSS (open-source software) stage, specific to open-source AI software, comprises six phases, derived from an amalgamation of the project's duration and the number of releases.

\textbf{`Aged'.} When paired with either \textit{Time Stage} or \textit{OSS Stage}, `Aged' signifies calculations that are predicated on the software component lifetime, rather than the entire duration of the project's existence.

\textbf{Days Difference} and \textbf{Age}. \textit{Days Difference} denotes the time span from the start of the project until a vulnerability is identified. Meanwhile, \textit{Age} signifies the time elapsed for a specific software component up to the point of vulnerability discovery.

\textbf{Classic Metrics.} In this context, classic metrics (`\textit{Classic}') encompass Code churn, Churn rate, and File size.

\subsection{Metrics}

\subsubsection{Code Ownership Metric}
We refine code ownership metrics based on previously works, emphasizing the relationship between software components and contributors within the entire code commit progression. Unlike conventional methods that isolate software components, we include data from vulnerable commits, utilising commit attributes to understand the impact of project locality on the code ownership metric. This approach not only examines the thresholds of ownership definitions but also quantifies ownership in both absolute and relative terms, associating it with the time/release metric to identify potential correlations.

In OSS projects, individual source code files are considered as distinct components, a practice supported by inherent and external factors \cite{linuxsource}. This segmentation aligns with the operations of version control systems of Git and SVN, which track changes at file level. This approach promotes modularity, adhering to the single responsibility principle, and is crucial for software maintainability. We further consider each commit as an individual contribution, acknowledging that attackers often introduce vulnerabilities through small changes. By using a timeframe for the project and defining temporal attributes for each software component, the study enables a dual perspective, offering both a holistic view and a detailed analysis of the project's development and security landscape.

\subsubsection{Time and Release Metric}
Previous works highlight the significance of time \cite{midha2012factors, wynn2004organizational} and release \cite{english2007floss, wu2007modeling, YANG2016102, pachauri2015incorporating, pachauri2019reliability} attributes in influencing OSS projects. Different from other project features, time and release metrics diverge from examining specific code content, inherent project traits, or its application contexts. This essentially means they don't scrutinize the software component and its commit content in a granular manner. Instead, they offer a broad, statistical analytical viewpoint, versatile enough to be applied across diverse project evaluations. The time and release metrics initially segment the project (or software component) into sequential phases from T1 to T5 based on its duration. Concurrently, factoring in the number of releases, the project (or software component) gets classified into six distinct performance scenarios. Although the six stages may not follow a distinct linear progression, they are evaluated and assigned numerical values based on their maturity stages. Time and Release metrics are summarized in Table \ref{tab:time_release_metric}.

\begin{table*}
    \centering
    \renewcommand{\arraystretch}{1.2}
    \caption{Time and Release Metrics}
    \label{tab:time_release_metric}
    \begin{tabular}{l|p{8cm}|p{8cm}}
    \toprule
    \textbf{Metrics} & \textbf{Description} & \textbf{Operationalization} \\
    \midrule
    \multicolumn{3}{l}{\textbf{Time Stage}} \\
    \midrule
    T1   & The given time period is in 0 to 7 days. & \( T_1 : 0 < T \leq 7 \text{ days} \)\\
    T2   & The given time period is in 7 days to 3 months. & \( T_2 : 7 \text{ days} < T \leq 3 \text{ months} \) \\
    T3   & The given time period is in 3 months to 9 months. & \( T_3 : 3 \text{ months} < T \leq 9 \text{ months} \) \\
    T4   & The given time period is in 9 months to 3 years. & \( T_4 : 9 \text{ months} < T \leq 3 \text{ years} \) \\
    T5   & The given time period is over 3 years. & \( T_5 : T \geq 3 \text{ years} \) \\
    \midrule
    \multicolumn{3}{l}{\textbf{OSS Stage}} \\
    \midrule
    (1) SI   & \textit{Success Initiation.} The project has at least one successful release in the development. & \(SI = N_{\text{release}} \geq 1\) \\
    (2) TI   & \textit{Tragedy Initiation.} The project has no release within the given time period is greater than 1 year that the project is abandoned. & \(TI = T_{\text{since last release}} > 1 \text{ year} \land N_{\text{release}} = 0\) \\
    (3) II   & \textit{Indeterminate Initiation.} The project has no release within the given time period is less than 1 year but shows significant developer activity. & \(II = 0 \leq T_{\text{since last release}} < 1 \text{ year} \land N_{\text{release}} = 0\) \\
    (4) IG   & \textit{Indeterminate Growth.} The project has less than 3 releases but the given time period is within 1 year or has produced 3 releases where the growth phase is less than 6 months. & \(IG = (N_{\text{release}} < 3 \land T_{\text{since last release}} < 1 \text{ year}) \lor (N_{\text{release}} = 3 \land T_{\text{growth phase}} < 6 \text{ months})\) \\
    (5) SG   & \textit{Success Growth.} The project has at least 3 meaningful releases where the growth phase is greater than 6 months. & \(SG = N_{\text{release}} \geq 3 \land T_{\text{growth phase}} > 6 \text{ months}\) \\
    (6) TG   & \textit{Tragedy Growth.} The project has 1 or 2 releases while the project's latest release has been 1 year since its last update. & \(TG = (N_{\text{release}} = 1 \lor N_{\text{release}} = 2) \land T_{\text{since last release}} > 1 \text{ year}\) \\
    \bottomrule
    \end{tabular}
\end{table*}

\section{Data Collection and Analysis}
\subsection{Data Collection and Processing}
We focuses on the open-source AI software, which are five prominent deep learning projects of TensorFlow\footnote{https://github.com/tensorflow/tensorflow}, Caffe\footnote{https://github.com/BVLC/caffe}, OpenCV\footnote{https://github.com/opencv/opencv}, Keras\footnote{https://github.com/keras-team/keras}, and PyTorch\footnote{https://github.com/pytorch/pytorch}. Our primary data sources are the NIST official vulnerability database NVD
and the GitHub repositories.

In our study, we combine data collection and processing to construct a detailed vulnerability dataset. With vulnerability information from NVD and GitHub repositories, we use APIs to gather CVE details and scrutinized GitHub security advisories for insights into commits and pull requests linked to security issues. The initial dataset, containing URLs, dates, and CVE IDs, is refined for consistency. The raw data is then structured using GitHub's REST APIs, extracting details from pull requests and commits, including source files and timestamps, with CVE numbers from the NVD provided severity scores. We analyse the complete commit history of each project with GitPython, allowing to calculate relevant metrics and perform a comprehensive analysis of the vulnerability dataset, focusing on software component details and metric derivation.

In the meantime, we use latest material from TensorFlow project as a benchmark for non-vulnerability. Recognised for its robust release management practices, we anticipate that its latest version boasts significant security measures. In summary, the collected dataset for the study covers 39,323 records, segregated into 27,075 entries marked as vulnerable and 12,248 as non-vulnerable. Out of the vulnerable entries, 904 were annotated with detailed CVE information, providing a rich dataset for subsequent analysis.

\subsection{Analysis Techniques}

\begin{figure}
    \centering
    \includegraphics[width=\linewidth]{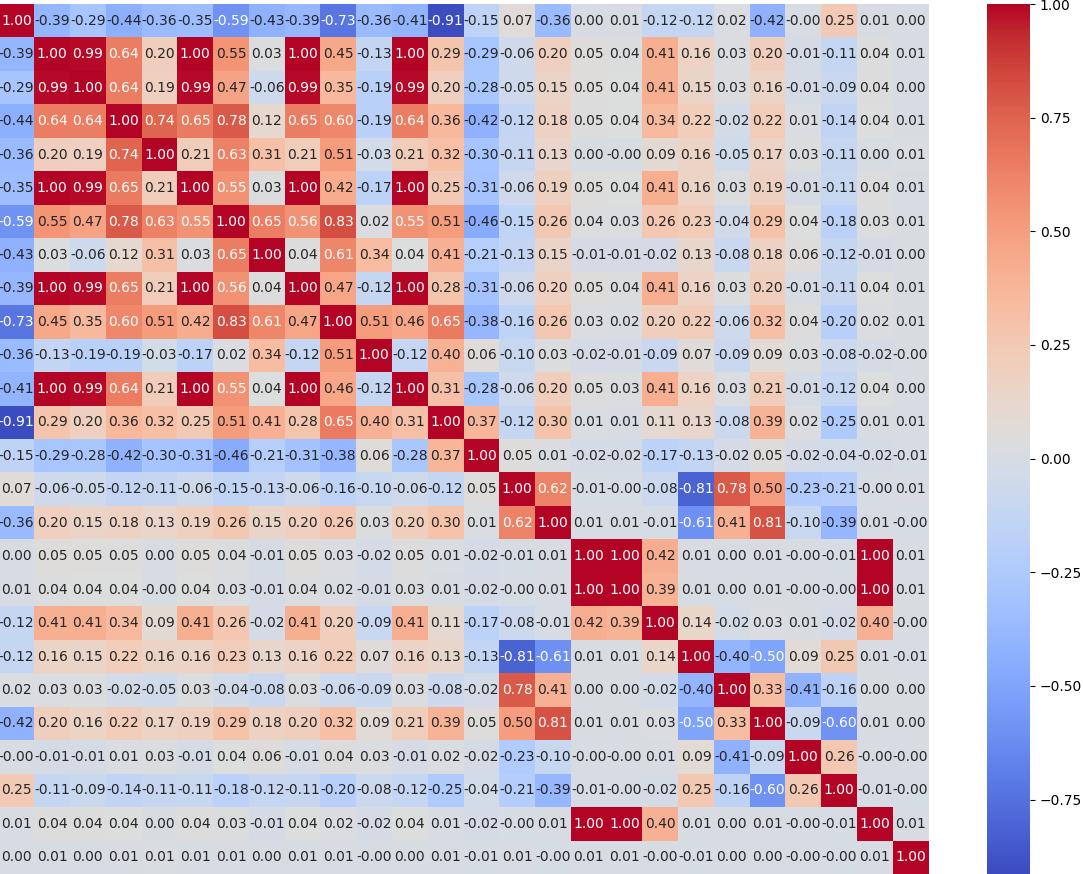}
    \caption{The example of Pearson Correlation Matrix Heatmap for metrics analysis (Pearson correlation value is within the range of [-1,1], representing the positive and negative correlationship between the metrics respectively; the metrics are from the set of \{\textit{Ownership, Num of contributor, Num of Minor, Per of Minor, Days Difference, Age, OSS Stage Aged, File Size, Code churn, Churn rate, etc.}\})}
    \label{fig:heatmap-example}
\end{figure}

In analyzing the impact of distortion factors on metric results, we employ a multi-faceted analytical approach. The core of our evaluation involves generating correlation heatmap matrices (i.e., Figure \ref{fig:heatmap-example}) under varying distortion factor values, with the Frobenius Norm \cite{mathworld_frobenius} as the foundational tool for identifying matrix discrepancies. To refine the analysis, we utilise Min-Max and exponential decay similarity scores \cite{min_max, heidary2019exponential}, which normalize and weight the discrepancies, ensuring finer nuances are captured. In addition to this, cosine similarity and Kolmogorov–Smirnov statistics provide insights into matrix alignment and distributional deviations, enhancing our ability to identify subtle pattern shifts. Further validation is done through the Mantel test \cite{mantel_1967}, examining the correlation between two distance matrices. This layered approach, integrating multiple analytical techniques, offers a comprehensive yet computationally demanding method to assess and validate the influence of distortion factors on our metric, without necessarily assuming data normality.

Later on, we assess the relationship between the newly proposed metric and software vulnerability, utilising three correlation methods: Pearson \cite{pearson1895}, Spearman \cite{spearman1904}, and Kendall \cite{kendall1938}. These methods, each generating correlation coefficients between -1 and 1, helped us quantify the strength and nature of variable relationships. Pearson's method, ideal for linear relationships under normal distribution, contrasted with Spearman's and Kendall's non-parametric approaches, is suitable for data rankings and non-normal distributions. Our analysis begins with an exploration of the connection between software vulnerability and code ownership, enhanced by a set of supplementary metrics. Incorporating a non-vulnerable dataset has facilitated a comparative study, enhancing our understanding of code ownership metrics in different software contexts. Furthermore, the relationship between our metric, CVE severity, and software release information is also scrutinized. 

To address the potential confounding effects of software component attributes such as churn rate and file size, Multiple Linear regression \cite{galton1886, freedman2009} is incorporated in our analysis. This regression model is instrumental in dissecting the influence of multiple variables on a targeted outcome, maintaining the constancy of other factors. Our primary aim is to observe changes in adjusted \( R^2 \), F-statistic and Coefficient values, which indicate the extent to which independent variables explain the variance in the dependent variable. This methodology provides insights related to the impact changes in one metric when controlled by another, offering a deeper understanding of metric robustness. To alleviate the risks of multicollinearity and overfitting, we carefully incorporate only one metric category as a control in each iteration of the model.

\section{Results}
\subsection{Potential Distortion Factor Check
\label{cha:distortion_check}}

To evaluate the influence of vulnerability file/commit frequency on our chosen metric, we juxtapose the metric's computational outcomes with a non-vulnerable dataset. We then quantify the heatmap, constructing diverse metric result correlation matrices by modulating the vulnerable to non-vulnerable ratio from 1:10 (10\%) to 1:1 (100\%). The average scores derived from Min-Max Scaling and Exponential Decay are 0.7127 and 0.4699, respectively with the fact that the p-values from the Mantel test across varying scales gravitate closely to 1. This underscores that alterations in vulnerability occurence proportion do not critically perturb the interrelation amidst matrices.

Promptly, this study embarks on an investigation into the potential effects of varying threshold definitions for minor contributors on metric measurements. Utilizing distinct thresholds (5\%, 10\%, 20\%, 50\%), we generate corresponding metric result correlation heatmaps. The cosine similarity exhibits a noticeable divergence between the 5\% and 50\% thresholds as around 0.58 to 0.91. However, the 10\% threshold remains in close alignment with the other three at a high level, which is between 0.72 to 0.91. Furthermore, by converting the matrices into one-dimensional arrays via reshaping or vectorization and employing the two-sample K-S test to assess the distributions of these arrays, the findings do not demonstrate a significant deviation in distribution (i.e., the p-value substantially exceeds 0.05). This suggests a consistent matrix distribution regardless of the applied thresholds, which indicates that 10\% is the optimal threshold to define the minor contributor.

In the next step, we delve into the influence of locality clustering on the metric evaluation. Leveraging the software component as defined by the metric, we synthesise two correlation heatmap matrices corresponding to file and group components (with group components clustered by pull request or commit origin). Their interrelation is assessed using the Mantel test. As analysed in the experiment, the correlation between the matrices is strikingly high at 0.822, signifying the congruence of the two datasets. This underscores that the software component's locality does not impact any significant effect on the metric measurements. 

From the comprehensive discussions and analyses presented, it is evident that the efficacy of metric measurements remains consistent, uninfluenced by factors such as vulnerability occurrence rate, minor contributor threshold definition, and software locality. Concurrently, this reinforces the validity of Hypothesis 2 and 3, suggesting that the susceptibility of software components to vulnerabilities is decoupled from their vulnerability occurrence rate. Moreover, software components behave independently, irrespective of their positioning within the project's framework.

\subsection{Correlation Analysis}

Initially, we attempt to see a direct association between metrics and vulnerability. While prior research, shown by the work of Bird et al., categorized datasets based on pre- and post-release vulnerabilities, there is no effort to integrate vulnerable and non-vulnerable data to distinctly assess the linkage between metrics and vulnerability. In our approach, we merge the vulnerable and non-vulnerable results in a 1:1 ratio, building on the premise discussed earlier that the proportion of vulnerability does not sway data measurement outcomes (see Sec. \ref{cha:distortion_check}). Given the diverse result distributions across different metrics, we evaluate the Pearson, Spearman, and Kendall correlation scores. As delineated in Table \ref{tab:direct_correlation_metrics_vulnerability}, it's evident that Time metrics (specifically, \textit{Days difference} and \textit{Age}, score at -0.81 and -0.61 respectively) possess a potent negative correlation with vulnerability (\textbf{RQ1}). This implies that as time progresses, vulnerability likelihood diminishes. Conversely, other metrics do not demonstrate any notable correlation with vulnerability, irrespective of the time frame or without classification.

\begin{table}[]
\caption{Correlation analysis of metrics with vulnerability presence in a balanced dataset of vulnerable and non-vulnerable sources}
\label{tab:direct_correlation_metrics_vulnerability}
\begin{tabular}{l|l|l|l}
\toprule
\textbf{Metrics}         & \multicolumn{3}{l}{\textbf{Is Defective}}               \\
                         & \textbf{Pearson} & \textbf{Spearman} & \textbf{Kendall} \\
                         \midrule
Ownership                & -0.12            & -0.1              & -0.08            \\
Num of contributor       & 0.16             & 0.22              & 0.19             \\
Num of Minor        & 0.16             & 0.23              & 0.2              \\
Per of Minor        & 0.23             & 0.23              & 0.2              \\
Days Difference          & -0.81            & -0.92             & -0.82            \\
Age                      & -0.61            & -0.6              & -0.49            \\
Oss Stage Aged   & 0.25             & 0.28              & 0.27             \\
File Size                & 0.14             & 0.4               & 0.33             \\
Code churn               & 0.01             & 0.36              & 0.3              \\
Churn rate               & -0.01            & 0                 & 0     \\ 
\bottomrule
\end{tabular}
\end{table}

After a comprehensive examination of the relationship between metrics and vulnerability, and after analyzing the correlation matrix heatmap for each metric, we segment vulnerability results based on the newly introduced metric '\textit{Time Stage Aged}'. We then investigate the correlation between the vulnerability classified by time attribute and metrics. Additionally, we juxtapose our pre-/post-release classification methods with those from prior studies. As illustrated in Table \ref{tab:correlation_metrics_time}, when vulnerabilities are categorized by the '\textit{Time Stage Aged}', a distinct correlation with code ownership emerges. Within the table, \textit{ownership} displays a pronounced negative correlation, registering a Spearman coefficient of -0.62. Meanwhile, metrics such as '\textit{Num of Contributor}', '\textit{Num of Minor}', and '\textit{Per of Minor}' exhibit positive correlations with vulnerability, with Spearman scores of 0.7, 0.64, and a Pearson score of 0.64, respectively (\textbf{RQ1}). While classic metrics also manifest some degree of correlation, they lag behind the correlation strength of code ownership metrics, with the peak value recorded at 0.49 (Spearman) (\textbf{RQ2}). Concurrently, our analysis of the correlation between metrics and pre-/post-release classifications reveals that neither code ownership nor classic metrics hold significant correlations, with their absolute values all falling below 0.1 (\textbf{RQ2}). From the aforementioned analysis, it's evident that code ownership metrics possess a marked correlation with vulnerability when vulnerabilities are classified by \textit{Time Stage}. This also underscores that in the realm of Open-source AI software, classifying vulnerabilities by \textit{Time Stage} is more pertinent and efficacious than a rudimentary pre-/post-release dichotomy.

\begin{table}[]
\caption{Correlation results between metrics and vulnerability classified by Time Stage/Pre-release(PR)/Post-release(PO)}
\label{tab:correlation_metrics_time}
\begin{tabular}{l|lll|l|l}
\toprule
\textbf{Metrics} & \multicolumn{3}{l|}{\textbf{Time Stage Aged}}    & \textbf{PR} & \textbf{PO} \\
                 & \textbf{P} & \textbf{S} & \textbf{K} & \textbf{S} & \textbf{S} \\ \midrule
Ownership                         & -0.73            & -0.62             & -0.5             & 0.03              & -0.03             \\
Num of contributor                & 0.38             & 0.7               & 0.58             & -0.06             & 0.07              \\
Num of Minor                 & 0.36             & 0.64              & 0.53             & -0.06             & 0.07              \\
Per of Minor                 & 0.64             & 0.63              & 0.52             & -0.06             & 0.07              \\
Days Difference                   & 0.22             & 0.32              & 0.25             & -0.69             & 0.72              \\
Age                               & 0.81             & 0.96              & 0.86             & -0.35             & 0.37              \\
Oss Stage Aged            & -0.58            & -0.59             & -0.51            & -0.11             & -0.11             \\
File Size                         & 0.17             & 0.25              & 0.19             & 0.06              & 0.05              \\
Code churn                        & 0.02             & 0.49              & 0.38             & -0.07             & 0.07              \\
Churn rate                        & 0.09             & 0.45              & 0.36             & 0.01              & 0.03              \\ 
\bottomrule    
\end{tabular}
\par
\centering
\vspace{1pt}
\textit{Note:} P, S, K represent Pearson, Spearman, and Kendall correlation respectively
\end{table}

In an intriguing observation detailed in Table \ref{tab:correlation_metrics_time}, both the code ownership metric and the classic metric display noteworthy correlations with vulnerabilities. Moreover, the `\textit{OSS Stage Aged}' metric also demonstrates a significant negative correlation with vulnerability. To delve deeper into this phenomenon, we enriched our dataset with CVE severity data and integrated it with release information. Subsequently, we introduce `\textit{Release Amounts}', a metric signifying the count of releases impacted by vulnerabilities within a defined timeframe. As presented in Table \ref{tab:correlation_metrics_cve}, for CVE severity, both `\textit{Days Difference}' and `\textit{Release Amounts}' exhibit positive correlations, boasting Pearson correlation coefficients of 0.45 and 0.43, respectively. Concurrently, pre-/post-release classifications display correlations of -0.41 and 0.4, respectively (\textbf{RQ3)}. These findings suggest that with the elongation of a project repository's lifespan and an escalation in the number of releases, there is a corresponding increase in vulnerability severity. Furthermore, vulnerabilities emerging post-release tend to be more detrimental compared to those identified pre-release.

\begin{table}[]
\caption{Correlation results between metrics and vulnerability with CVE severity}
\label{tab:correlation_metrics_cve}
\begin{tabular}{l|lll}
\toprule
{\textbf{Metrics}} & \multicolumn{3}{l}{\textbf{CVE Severity}}              \\
                                  & \textbf{Pearson} & \textbf{Spearman} & \textbf{Kendall} \\ 
\midrule
Ownership                         & -0.03            & -0.07            & -0.05            \\
Num of contributor                & 0.11             & 0.08             & 0.06             \\
Num of Minor                 & 0.12             & 0.09             & 0.06             \\
Per of Minor                & 0.06             & 0.07             & 0.05             \\
Days Difference                   & 0.45             & 0.44             & 0.35             \\
Age                               & 0.25             & 0.26             & 0.19             \\
Oss Stage Aged            & -0.02            & -0.02            & -0.02            \\
File Size                         & 0.1              & 0.16             & 0.12             \\
Code churn                        & 0.12             & 0.14             & 0.1              \\
Churn rate                        & 0.06             & 0.01             & 0.01             \\
Is Pre-release                    & -0.41            & -0.32            & -0.28            \\
Is Post-release                   & 0.4              & 0.31             & 0.28             \\
Release Amounts                   & 0.43             & 0.43             & 0.34             \\
Release Amounts Aged              & 0.27             & 0.29             & 0.21             \\
\bottomrule
\end{tabular}
\end{table}

Finally, in our analysis, we employ multiple linear regression techniques to methodically evaluate interrelationships among designated indicators. As delineated in Table \ref{tab:metric_robustness_check}, we devise several statistical vulnerability models rooted in metrics showing significant correlation, incorporating distinct metric values and appraising the efficacy of their respective combinations (controlled by different metrics). Notably, the amalgamation of `\textit{Oss Stage Aged}' with `\textit{Per of Minor}' manifested a 33\% enhancement in the adjusted R-squared value. Contrarily, the juxtaposition of other metrics yielded negligible variations in adjusted R-squared (\textbf{RQ2}). These empirical observations underscore the predominant independence of the metrics in consideration. While a subtle association between `\textit{Per of Minor}' and `\textit{Oss Stage Aged}' is discernible, its significance remains marginal. This implies that while most metrics predominantly operate independently, a specific interplay between `\textit{Oss Stage Aged}' and `\textit{Per of Minor}' might, under certain conditions, have a more pronounced influence on the relationship between software components and vulnerabilities. Nevertheless, this observation stands as a hypothesis that requires further investigation in the future.

\begin{table}[]
\caption{Metric robustness check}
\label{tab:metric_robustness_check}
\centering
\begin{tabular}{l|lll}
\toprule
\textbf{Model}                                                         & \textbf{Adj. R\textsuperscript{2}} & \textbf{F-statistic} & \textbf{Coef.} \\ 
\midrule
\multicolumn{4}{l}{\textbf{Is Defective}}                                                                                                                                            \\ 
\midrule
Days Difference                                          & 0.659            & 4.58E+04             & -0.0005              \\
Days Difference + Classic          & 0.665            & 1.57E+04             & -0.0005              \\
Age                                                      & 0.367            & 1.37E+04             & -0.0003              \\
Age + Classic                      & 0.388            & 5020                 & -0.0003              \\ 
\midrule
\multicolumn{4}{l}{\textbf{Time Stage Aged Numeric}}                                                                                                                                 \\ 
\midrule
Per of Minor                                        & 0.413            & 8099                 & 1.9658               \\
Per of Minor + Classic & 0.416            & 2723                 & 2.0164               \\
Oss Stage Aged                                   & 0.332            & 5715                 & -0.5741              \\
Oss Stage Aged + Classic   & 0.347            & 2033                 & -0.5628              \\
Per of Minor + Oss Stage Aged & 0.553            & 7093                 & 1.1372               \\ 
\midrule
\multicolumn{4}{l}{\textbf{CVE Severity}}                                                                                                                                            \\ 
\midrule
Days Difference                                          & 0.202            & 222.7                & 0.0021               \\
Days Difference + Classic          & 0.203            & 74.64                & 0.002                \\
Days Difference + Minor                    & 0.202            & 111.4                & 0.002                \\ 
\bottomrule
\end{tabular}
\end{table}

\subsection{Discussion}

The findings of this study underscore a significant relationship between code ownership metrics and vulnerabilities within open-source AI software projects. Moreover, the newly introduced code ownership and time metrics present a more potent correlation with vulnerabilities than classic metrics, thereby addressing the posed research questions:

\textbf{RQ1:} Our analysis reveals a pronounced correlation between code ownership metrics and open-source AI software vulnerabilities. Both `\textit{Num of Minor}' (calculated from absolute values) and `\textit{Per of Minor}' (based on the proportion of minor contributors) solidify this relationship. Code ownership emerges as a consistent trait across multiple open-source AI software projects, serving as a reliable measure of developer activity levels. The newly introduced time metrics, notably `\textit{Days Difference}' and `\textit{Age}', further reinforce this robust relationship. Notably, there is a discernible correlation between code ownership metrics and the `\textit{Time Stage Aged}' metric, suggesting that the proposed five-tier time stage model aptly encapsulates the project's evolutionary patterns over distinct durations.

\textbf{RQ2:} In terms of metric definitions, the `\textit{Time Stage Aged}' methodology for classifying vulnerabilities appears more aligned with open-source AI software projects compared to the pre-/post-release approach by Bird et al. Moreover, code ownership metrics consistently outperform classic metrics. It's also pertinent to note that software components, defined as individual source files, retain their autonomy, ensuring they remain uninfluenced by potential intra-project locality attributes. The novel metrics introduced in this study further exhibit independence, maintaining their efficacy irrespective of the thresholds defined for minor contributors or vulnerability occurrences.

\textbf{RQ3:} Prior analyses suggest that the severity of vulnerabilities amplifies as the project's lifespan extends. This severity is directly proportional to both the number of releases and post-releases and inversely related to pre-releases. Concurrently, an uptrend in minor contributors is observed as time advances, while the ownership metric shows a marked decrement. Summarizing, for effective management of open-source AI software projects, post-release phases warrant enhanced scrutiny. Additionally, prolonged software components with a burgeoning number of minor contributors demand rigorous oversight and governance.

\section{Threat to Validity}

This work demonstrates that newly proposed code ownership metrics combined with time/release attributes effectively reveals correlation on vulnerabilities in open-source AI software. Despite its effectiveness, there are some limitations for future research.

An extensive literature review indicates that dependency modification is a frequent attack strategy in open-source software projects. Dependencies are crucial to these projects, with Plate et al. \cite{plate2015impact} noting the risk of using outdated library fragments. Pashchenko et al. \cite{pashchenko2018vulnerable} found that non-direct, multi-level dependencies in these projects are often overlooked. Maintenance and release timing of dependencies, a factor often ignored, can be critical for security. Ponta et al. \cite{ponta2018beyond}, along with other studies \cite{kula2018developers, mostafa2017experience}, highlight the challenges in dependency management and its impact on vulnerability. This suggests a link between dependency handling, developer activity, and code ownership.

Meanwhile, the considerations emphasize the importance of a comprehensive approach in metric design and evaluation to fully capture software development and vulnerability aspects:

\textbf{Project Attribute Limitations.} Our study, centered on the link between developer activity and software vulnerabilities, may overlook influences like programming languages and license types \cite{colazo2009impact}. Foucault's work \cite{foucault2014ownership} shows varied major contributor distributions across projects. While minor contributor patterns and bug densities appear consistent, this might be unique to open-source AI projects. We assume software modules function independently of project elements, a premise needing further exploration.

\textbf{Data Quality \& Generalizability.} Based on five key open-source AI projects, our findings might not represent all such software. We used NVD's CVE severity scores, which could affect metric validity. Our reference for a non-vulnerable dataset, the latest TensorFlow version, may not be the most suitable choice.

\textbf{Metric Completeness.} The exclusion of complexity analysis in diverse programming languages may affect accuracy, a factor often included in traditional metrics \cite{bird2011donttouch}. Our approach to locality is basic and might not capture intricate component interactions within projects. Additionally, not considering the varying roles of contributors, a point emphasized by Bird et al. \cite{bird2011donttouch}, could overlook important elements influencing the effectiveness of these metrics.

\section{Conclusion}

In this work, we thoroughly emphasise the significance and effectiveness of code ownership in securing open-source AI software projects, presenting a novel perspective for improving security maintenance and evaluation. It demonstrates a significant correlation, wherein stronger ownership aligns with fewer defects. There is a notable positive correlation between the number and proportion of minor contributors and vulnerability. Innovative time-based metrics is introduced, highlighting the project's time span, the lifespans of individual software components, and the number of impacted releases. These temporal metrics proficiently delineate various vulnerability stages within a project, providing a solid base for code ownership assessment. Our work suggests that the time-release categorization accurately reflects a project's maturity and the severity of its vulnerabilities. Consequently, recommendations have been proposed for project managers to closely monitor projects with lengthy lifespans and distinct ownership patterns and to thoroughly examine those components with minimal ownership. In addition to these findings, a Python-based command-line tool is released to facilitate the assessment of these metrics in practical scenarios.

\balance
\bibliographystyle{ACM-Reference-Format}
\bibliography{sample-base}

\end{document}